\documentclass[aps,prl,reprint,superscriptaddress,floatfix]{revtex4-1}
\usepackage{graphicx}
\usepackage{color}
\usepackage{multirow}
\usepackage{amsmath}
\usepackage{mathtools}
\usepackage{bm}
\usepackage{epstopdf}
\usepackage{siunitx}
\DeclareMathOperator{\sech}{sech}
\renewcommand{\section}{\paragraph}

\begin{document}
\title{Interfacial contributions to spin-orbit torque and magnetoresistance in ferromagnet/heavy-metal bilayers}

\author{K. D. Belashchenko}
\affiliation{Department of Physics and Astronomy and Nebraska Center for Materials and Nanoscience, University of Nebraska-Lincoln, Lincoln, Nebraska 68588, USA}

\author{Alexey A. Kovalev}
\affiliation{Department of Physics and Astronomy and Nebraska Center for Materials and Nanoscience, University of Nebraska-Lincoln, Lincoln, Nebraska 68588, USA}

\author{M. van Schilfgaarde}
\affiliation{Department of Physics, King's College London, Strand, London WC2R 2LS, United Kingdom}

\date{\today}

\begin{abstract}
The thickness dependence of spin-orbit torque and magnetoresistance in ferromagnet/heavy-metal bilayers is studied using the first-principles non-equilibrium Green's function formalism combined with the Anderson disorder model. A systematic expansion in orthogonal vector spherical harmonics is used for the angular dependence of the torque. The damping-like torque in Co/Pt and Co/Au bilayers  can be described as a sum of the spin-Hall contribution, which increases with thickness in agreement with the spin-diffusion model, and a comparable interfacial contribution. The magnetoconductance in the plane perpendicular to the current in Co/Pt bilayers is of the order of a conductance quantum per interfacial atom, exceeding the prediction of the spin-Hall  model by more than an order of magnitude. This suggests that the ``spin-Hall magnetoresistance,'' similarly to the damping-like torque, has a large interfacial contribution unrelated to the spin-Hall effect.
\end{abstract}

\maketitle

\section{Introduction.}
Magnetization in nanoelectronic devices can be manipulated by angular momentum transfer \cite{SLONCZEWSKI1996L1,Berger:PRB1996}. Spin-orbit torque (SOT) \cite{MihaiMiron.GaudinNM2010,Liu.Moriyama.eaPRL2011,Manchon} is a manifestation of such transfer driven by spin-orbit coupling in ferromagnet/heavy metal (FM/HM) bilayers, which is also responsible for anisotropic transport properties such as spin-Hall magnetoresistance (SMR) \cite{PhysRevLett.110.206601,Chen2016}. SOT and SMR involve complex processes both in the bulk and at the interface \cite{Manchon,Amin.Stiles.PRB2016,Amin.Stiles.PRB2016a,PhysRevLett.117.207204,PhysRevLett.122.077201,Chen2016}.

SMR is usually explained by interfacial absorption of the spin-Hall current incident from the bulk of the heavy metal \cite{SMR-Chen}. It is also common to attribute damping-like SOT to the spin-Hall effect \cite{Sinova.ValenzuelaRMP2015} originating in the bulk of the heavy metal \cite{Liu.Moriyama.eaPRL2011,Liu.PaiS2012,Liu.LeePRL2012} and field-like SOT to the inverse spin-galvanic effect \cite{AronovGeller1989JETPL,EDELSTEIN1990233,Ganichev} at the interface \cite{Chernyshov.OverbyNP2009,MihaiMiron.GaudinNM2010,Miron.Garello2011,Manchon.KooNM2015,Fan.CelikNC2014}. However, damping-like SOT can also be generated at the interface  \cite{Pesin2012,Qaiumzadeh.DuinePRB2015,Ado.TretiakovPRB2017,Amin.Stiles.PRB2016,Amin.Stiles.PRB2016a} without any spin-polarized current incident from the bulk of the HM layer.
Besides, the layers in FM/HM bilayers are usually about a nanometer thick or even less. Thus, the distinction between interfacial and bulk contributions to SOT is not well defined, which prompts a fully quantum-mechanical treatment of the whole device \cite{Belashchenko2019}. This point of view is supported by recent experiments suggesting a competition between interfacial and bulk contributions to damping-like SOT \cite{Kim.Sinha.eaNM2013,Ramaswamy.Qiu.eaAPL2016,Torrejon.Kim.eaNC2014} and by \textit{ab-initio} calculations hinting at the importance of interfacial contributions \cite{PhysRevLett.116.196602,Belashchenko2019}. Most \textit{ab-initio} calculations of SOT have been performed using the linear response method with phenomenological broadening for the Green's functions \cite{Freimuth.BluegelPRB2014} and did not reveal the interfacial contribution to damping-like SOT \cite{PhysRevB.97.224426}.

In this paper, we employ the non-equilibrium Green's function (NEGF) approach \cite{PhysRevB.71.195422,Nikolic2018} within the tight-binding linear muffin-tin orbital (LMTO) method \cite{LMTO,Questaal} for \emph{ab-initio} calculations of SOT and SMR in magnetic multilayered systems with explicit treatment of disorder \cite{Belashchenko2019}. To identify the effective interfacial and bulk contributions, we study the thickness dependence of SOT, whose angular dependence is represented by an expansion in orthonormal vector spherical harmonics. For the leading damping-like SOT, we find a contribution increasing with thickness in agreement with the spin-diffusion model and a comparable interfacial contribution that survives in the limit of zero thickness. We further identify the interfacial contribution to magnetoresistance which exceeds the expected spin-Hall contribution by more than an order of magnitude.

\section{Thickness dependence of SOT.}

The technical details of the \textit{ab initio} NEGF calculations are similar to Ref.~\onlinecite{Belashchenko2019}, except that here we represent the angular dependence of SOT using the complete and orthonormal basis of vector spherical harmonics, as described in the Appendix. Only the Fermi-surface contribution is considered here, because the Fermi-sea term is considerably smaller at room temperature \cite{Belashchenko2019}. We employ the Anderson disorder model with a uniformly distributed random potential $V_i$, $-V_m < V_i < V_m$, applied on each lattice site $i$ with an amplitude $V_m=0.77$ or 1.09 eV. The SOT is calculated from the non-equilibrium spin density matrix on each atom \cite{Belashchenko2019}.

\begin{figure}[htb]
\includegraphics[width=0.9\columnwidth]{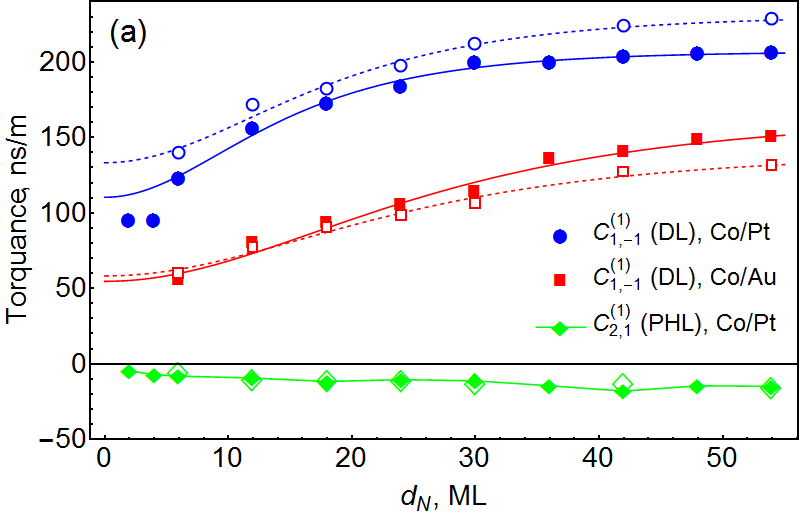}\vskip2ex
\includegraphics[width=0.9\columnwidth]{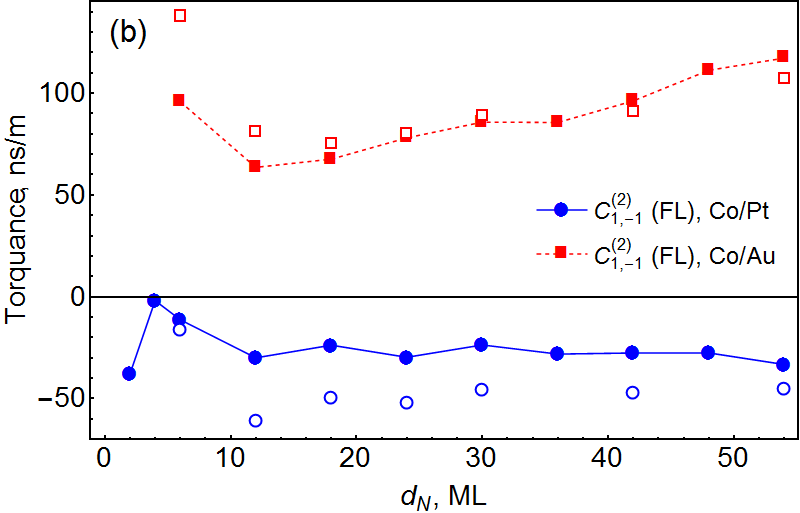}
\caption{Dependence of the SOT coefficients on the thickness $d_N$ of the heavy-metal layer in Co/Pt and Co/Au bilayers. (a) Dampinglike (DL) $C^{(1)}_{1,-1}$ and planar-Hall-like (PHL) $C^{(1)}_{2,1}$ coefficients. (b) Fieldlike (FL) coefficient $C^{(2)}_{1,-1}$. Blue (red) circles (squares): $C^{(1)}_{1,-1}$ or $C^{(2)}_{1,-1}$ for Co/Pt (Co/Au); green lines and diamonds: $C^{(1)}_{2,1}$ for Co/Pt. Filled (empty) symbols: $V_m=1.09$ eV (0.77 eV). Blue (red) lines in panel (a): fits of the data for Co/Pt (Co/Au) at $V_m=1.09$ eV (see text).}
\label{SOTvsD}
\end{figure}

Figure \ref{SOTvsD} shows the dependence of SOT in Co/Pt and Co/Au bilayers on the thickness of the heavy-metal layer $d_N$ measured in monolayers (ML). The damping-like SOT coefficient $C^{(1)}_{1,-1}$ is well described by the function
\begin{equation}
C^{(1)}_{1,-1}=\tau_0 + \tau_\mathrm{SH}\left[1-\sech(d_N/l_{sf})\right]
\label{intbulk}
\end{equation}
with the parameters listed in Table \ref{tab:fit}, where $\tau_0$ represents the thickness-independent interfacial contribution to SOT and $\tau_\mathrm{SH}$ the conventional spin-Hall-generated part \cite{Liu.Moriyama.eaPRL2011}. However, the fitting (\ref{intbulk}) should be approached with care, because it assumes a geometrical interface between homogeneous bulk regions and ignores thickness-dependent perturbations and finite-size effects in the electronic structure of the bilayer.

Importantly, we find that $\tau_0$, which can appear due to interface scattering \cite{Amin.Stiles.PRB2016,Amin.Stiles.PRB2016a}, is comparable with $\tau_\mathrm{SH}$ in both systems and is especially large in Co/Pt. The value of $\tau_\mathrm{SH}$ is similar in Co/Pt and Co/Au, while the spin-diffusion length is larger in Co/Au.

\begin{table}[htb]
\begin{tabular}{|c|c|c|c|c|c|c|}
\hline
     FM/HM & $V_m$, eV & $\bar\rho$, \si{\micro\ohm\centi\meter} & $\tau_0$, ns/m & $\tau_\mathrm{SH}$, ns/m & $l_{sf}$, nm & $\theta_\mathrm{SH}$\\
     \hline
     \multirow{2}{*}{Co/Pt} & 1.09  &   27.2    & 110.3 & 96.5 & 1.94 & 0.027\\
     \cline{2-7}
                            & 0.77  &   17.7    & 133.2 & 97.5 & 2.43 & 0.018\\
     \hline
     \multirow{2}{*}{Co/Au} & 1.09  &   6.4     & 54.6 & 108.9 & 3.61 & 0.007\\
     \cline{2-7}
                            & 0.77  &   4.1     & 58.2 & 81.1  & 3.38 & 0.003\\
     \hline
\end{tabular}
\caption{Parameters of the fits [Eq.\ (\ref{intbulk})] for the thickness dependence of the damping-like SOT coefficient $C^{(1)}_{1,-1}$. The effective resistivities $\bar\rho$ are given for the Co(4 ML)/HM(12 ML) bilayers.}
\label{tab:fit}
\end{table}

The influence of disorder strength on the damping-like SOT can be seen by comparing the results for $V_m=1.09$ and 0.77 eV in Fig.\ \ref{SOTvsD}a (filled and empty symbols, respectively) and Table \ref{tab:fit}. The behavior is different in Co/Pt and Co/Au bilayers. In Co/Pt, increasing $V_m$ from 0.77 to 1.09 eV reduces $l_{sf}$ and $\tau_0$, both by about 20\%, while $\tau_\mathrm{SH}$ remains unchanged. This suggests that the spin-Hall contribution to damping-like SOT in Co/Pt is dominated by a combination of intrinsic and side-jump mechanisms. In contrast, stronger disorder in Co/Au leads to a 34\% increase in $\tau_\mathrm{SH}$ while the changes in $l_{sf}$ and $\tau_0$ are within the margin of error. The dependence of $\tau_\mathrm{SH}$ on $V_m$ in Co/Au suggests that the skew-scattering contribution, which is proportional to the relaxation time, is not negligible and negative in this system. The suppression of the skew scattering contribution in Co/Pt it likely due to the much larger resistivity in this system compared to Co/Au.

Interestingly, $l_{sf}$ does not increase with increasing conductivity in Co/Au, suggesting that the Dyakonov-Perel mechanism of spin relaxation, which was shown to be dominant in some supported metallic films \cite{Long2016}, can survive even in the presence of a proximate ferromagnetic layer. This mechanism is expected to be more important in Co/Au compared to Co/Pt, because the magnetic proximity effect in Au is much weaker than in Pt.

We have previously found  \cite{Belashchenko2019} that spin-orbit coupling on the Co atoms has no effect on the total damping-like SOT in the Co(6 ML)/Pt(6 ML) bilayer. The same statement holds for Co(4 ML)/Pt(30 ML) and Co(4 ML)/Au(30 ML) bilayers, suggesting that the spin-Hall effect in Co \cite{AminSHEFM} does not contribute materially to damping-like SOT.

Table \ref{tab:fit} also lists the effective spin-Hall angle in the heavy metal estimated as
\begin{equation}
    \theta_\mathrm{SH}=\sqrt{\frac{3}{8\pi}}\frac{2e}{\hbar}\tau_\mathrm{SH}\rho\frac{M}{A}
\end{equation}
where $M/A$ is the total magnetic moment per area of the film, and the numerical factor comes from Eq.~(\ref{dlvsh}). Small $\theta_\mathrm{SH}$ in Co/Au is due to the low resistivity of Au compared to Pt. The values for Co/Pt are smaller compared to the experimental measurements; for example, $\theta_\mathrm{SH}\approx0.06$ was reported in Ref.~\onlinecite{Liu.Moriyama.eaPRL2011}. However, including the interfacial contribution $\tau_0$ would result in $\theta_\mathrm{SH}\approx0.06$, which may increase further in more resistive films. We also note that the spin-Hall conductivity estimated from $\tau_\mathrm{SH}$ assuming the spin-Hall current is fully absorbed by the magnetization is similar to the result of the Berry-phase calculation at zero temperature \cite{SHE-Pt}.

The planar-Hall-like term $C^{(1)}_{2,1}$ in Co/Pt is roughly proportional to the leading damping-like coefficient $C^{(1)}_{1,-1}$ and amounts to about $-7$\% of it. This term can be isolated experimentally by measuring current-induced magnetization damping in the $xz$ plane \cite{Safranski.Montoya.eaAe2017,Belashchenko2019}. It is seen from Fig.\ \ref{SOTvsD}a that disorder strength has little influence on $C^{(1)}_{2,1}$, consistent with its attribution to vertex corrections \cite{Belashchenko2019}. In Co/Au the planar-Hall-like term is positive, does not exceed 5 ns/m, and is not shown in Fig.\ \ref{SOTvsD}a.

The thickness dependence of the field-like SOT $C^{(2)}_{1,-1}$ is shown in Fig.\ \ref{SOTvsD}b. This term exhibits sharp variations at small thicknesses, which likely reflect the sensitivity of the underlying mechanism to the electronic structure near the interface. At larger thicknesses the field-like SOT converges to a moderate value in Co/Pt, but in Co/Au it is comparable to the damping-like SOT and keeps growing at large thicknesses. This growth may indicate that the field-like SOT in Co/Au has a contribution that is associated with the spin-Hall effect in Au.

The effect of disorder strength on the field-like SOT is also different in Co/Pt and Co/Au bilayers. While stronger disorder tends to suppress $C^{(2)}_{1,-1}$ in Co/Pt, which is consistent with inverse spin galvanic effect (ISGE), it has hardly any effect on it in Co/Au. This feature is due to the cancellation of disorder-induced trends in different bands. Indeed, if  spin-orbit coupling is turned off on the Co atoms, the field-like $C^{(2)}_{1,-1}$ coefficient at $d_N=30$ ML increases from 86 to 150 ns/m at $V_m=1.09$ eV, and from 89 to 226 ns/m at 0.77 eV. Thus, the contributions to the field-like torque from spin-orbit coupling on Co and Pt atoms have opposite signs and both decrease with increasing disorder, as expected for ISGE.

\section{Magnetoresistance.}

The term \emph{spin-Hall magnetoresistance} (SMR) refers to the reduction of the resistance of a FM/HM bilayer whose magnetization is aligned parallel to the interface and perpendicular to the current flow, i.e., along the $\hat y$ axis.
Measurements of the magnetoresistance in fully metallic bilayers \cite{Kim2016} including Co/Pt \cite{Kawaguchi2018} have been attributed to the spin-Hall mechanism.

The phenomenological theory of SMR \cite{SMR-Chen}, which was introduced for a bilayer with an insulating FM layer,
describes the effect of spin-dependent interfacial scattering of the spin-Hall current incident from the heavy-metal layer and predicts the $\cos^2\theta_y$ angular dependence of the magnetoresistance. However, such magnetoresistance can also arise due to a purely interfacial mechanism \cite{Grigoryan2014}. Metallic FM/HM bilayers also exhibit anisotropic magnetoresistance (AMR) with the $\cos^2\theta_x$ angular dependence contributed by the metallic FM layer. In addition, the dependence of the interfacial electronic structure on the orientation of the magnetization \cite{Bode2002} should result in an \emph{interfacial anomalous magnetoresistance} (IAMR) with the $\cos^2\theta_z$ angular dependence. IAMR is similar to the tunneling anisotropic magnetoresistance \cite{TAMR} but occurs in the current-in-plane geometry.

Because the three functions $\cos^2\theta_\alpha$ representing the angular dependence of SMR, AMR, and IAMR are linearly dependent, there are, in fact, only two independent parameters that can be extracted from experiment, and different mechanisms can not be uniquely separated from the angular dependence. On the other hand, only the SMR mechanism is due to the spin-Hall effect in the bulk of the heavy-metal layer, and it comes with a characteristic dependence on its thickness $d_N$ \cite{SMR-Chen}.

We introduce the reduced conductance $g(\hat m)=LG(\hat m)/w$ in the Ohmic limit (large $L$ and $w$), where $G(\hat m)$ is the conductance, $L$ the length, and $w$ the width of the bilayer, and write it as
\begin{equation}
    g(\hat m,d_N) = g_F(\hat m,d_N) + \sigma_N d_N
\end{equation}
where $\sigma_N$ is the conductivity and $d_N$ the thickness of the normal metal. The two terms on the right-hand side represent, respectively, the angular-dependent contribution of the FM layer (including the interface), and the angular-independent bulk contribution of the normal metal. The dependence of $g_F$ on $d_N$ can come both from the spin-Hall contribution and from Friedel oscillations and quantum-well-like effects in other mechanisms. Of course, the angular dependence of $g_F$ also includes the contribution of AMR.

Defining the difference $\Delta_{\mu\nu}g(d_N)=g(\hat\mu,d_N)-g(\hat\nu,d_N)$, where $\hat\mu$, $\hat\nu$ are some chosen orientations of $\hat m$, we have $\Delta_{\mu\nu}g(d_N) =\Delta_{\mu\nu}g_F(d_N)$. The quantity $\Delta_{y z}g/g$ is usually reported as SMR. Note that $\Delta_{\mu\nu}g$ is expected to saturate at large thicknesses. The prediction of the spin-Hall theory \cite{SMR-Chen,Kim2016} for the angular and thickness dependence of $\Delta_{y z}g$ can be compared with the results of \emph{ab initio} calculations.

We compute $g(\hat x)$, $g(\hat y)$ and $g(\hat z)$ for Co/Pt bilayers with a varying thickness $d_N$ of the Pt layer while keeping the Co layer 4 monolayers thick. For each value of $d_N$, we calculate the conductance $G$ for a fixed width $w$ and $L=60,90,\dots,240$ ML, taking several hundred disorder configurations for each $L$. The dataset for the given $d_N$ and $w$ is then fitted to the function $G(L)=(R_0+Lg/w)^{-1}$, which provides the value of $g$ for the given $\hat m$. Finally, we find $\Delta_{y z}g(d_N)$ and $\Delta_{y x}g(d_N)$. Most of the calculations were performed with $w=12$ ML, but for $d_N=4$, 12 ML we also considered $w=16$ ML to check convergence with respect to $w$.

\begin{figure}[htb]
\includegraphics[width=0.9\columnwidth]{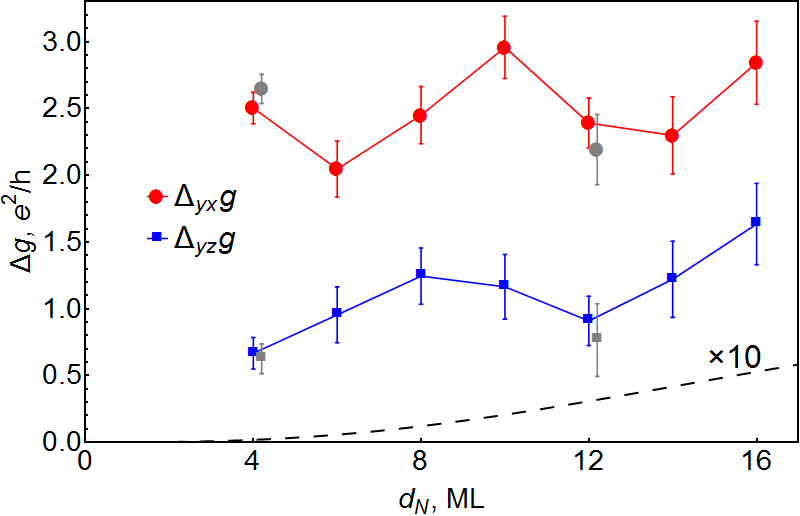}\hfil
\caption{Dependence of the magnetoconductances $\Delta_{yx}g$ (red symbols) and $\Delta_{yz}g$ (blue symbols) in Co/Pt bilayers on the thickness of the Pt layer calculated at $w=12$ ML. Grey symbols: convergence tests at $w=16$ ML. The disorder strength is $V_m=1.09$ eV. The error bars show the standard deviation. Dashed line: the prediction of Eq.~(\ref{SMR-model}) with the parameters for Co/Pt at $V_m=1.09$ eV from Table \ref{tab:fit}, scaled by a factor of 10.}
\label{SMR}
\end{figure}

The results of the calculations of $\Delta_{y z}g(d_N)$ and $\Delta_{y x}g(d_N)$, along with their standard deviations, are shown in Fig.~\ref{SMR}. The magnitude of $\Delta_{y x}g$, which is influenced by SMR and AMR but not IAMR, appears to be thickness-independent apart from relatively small oscillations. $\Delta_{yz}g$, which is influenced by SMR and IAMR but not AMR, is of the order of one conductance quantum, which is well above the margin of error, but its thickness dependence is obscured by the relatively large error bars.

The spin-Hall theory \cite{SMR-Chen,Kim2016} predicts the following spin-Hall contribution to $\Delta_{yz}g$, where we assume the magnetic layer is sufficiently thin so that the shunting effect can be neglected:
\begin{equation}
\Delta_{yz} g_\mathrm{SH}=\frac{\theta_\mathrm{SH}^2}{\bar{\rho}}\tanh^2(d_{N}/2l_{sf})\tanh(d_{N}/l_{sf}).
\label{SMR-model}
\end{equation}
The values of $\theta_\mathrm{SH}$, $l_{sf}$, and $\bar\rho$ obtained from SOT calculations can be found in Table \ref{tab:fit}. The dashed line in Fig.~\ref{SMR} shows the resulting $\Delta_{yz} g_\mathrm{SH}(d_N)$ scaled by a factor of 10. It is clear that the spin-Hall mechanism \cite{SMR-Chen,Kim2016} is too weak, by more than an order of magnitude, to account for the calculated magnetoconductance $\Delta_{yz} g\sim e^2/h$. The effect can, therefore, be entirely due to interfacial mechanisms.

The growth of $\Delta g_{yz}/g$ at small thicknesses \cite{Kim2016,Kawaguchi2018} is the main evidence in favor of the spin-Hall theory of magnetoresistance in FM/HM bilayers. However, this inference assumes that the interface is well-formed and has the same properties for all $d_N$. In a practical device, this assumption can fail once $d_N$ is reduced to a few monolayers, which is comparable to both the typical interfacial roughness and to the screening length of the metal. Thus, the region of linear growth of $\Delta g_{yz}/g$ could indicate the typical thickness at which the continuous and homogeneous heavy-metal film is formed during deposition. Interfacial roughness sets the natural scale for this thickness.

\section{Conclusions.}

The results of \emph{ab initio} calculations suggest that the damping-like SOT in Co/Pt and Co/Au bilayers has an interfacial contribution comparable to the spin-Hall effect in the heavy-metal layer. The magnetoconductance $\Delta_{yz}g$ for the Co/Pt bilayer is found to be of the order of a conductance quantum per interfacial atom, which exceeds the expected spin-Hall magnetoconductance by more than an order of magnitude and likely has an interfacial origin.

\begin{acknowledgments}

We thank Vivek Amin, Gerrit Bauer, Ilya Krivorotov, Farzad Mahfouzi, Branislav Nikoli\'c, and Mark Stiles for useful discussions. This work was supported by the National Science Foundation (NSF) through grant DMR-1609776 (K.B.) and the Nebraska Materials Research Science and Engineering Center (MRSEC) grant DMR-1420645 (K.B. and A.K.), the U.S. Department of Energy, Office of Science, Basic Energy Sciences award DE-SC0014189 (A.K.), and EPSRC CCP9 Flagship Project No.~EP/M011631/1 (M.v.S.). Calculations were performed utilizing the Holland Computing Center of the University of Nebraska, which receives support from the Nebraska Research Initiative.

\end{acknowledgments}

\section{Appendix: Expansion of SOT in vector spherical harmonics.}

Any vector field $\mathbf{V}(\hat n)$ defined on the unit sphere and tangential to it can be represented as a linear combination of orthonormal vector spherical harmonics (VSH) \cite{QED} $\mathbf{Y}_{lm}^{(\nu)}$ ($\nu=1,2$):
\begin{equation}
    \mathbf{V}(\hat n) = \sum_{lm\nu} A_{lm}^{(\nu)}\mathbf{Y}_{lm}^{(\nu)}(\hat n)
\end{equation}
where
\begin{align}
    \mathbf{Y}_{lm}^{(1)}(\hat n) &= \frac{r \nabla Y_{lm}(\hat n)}{\sqrt{l(l+1)}},\\
    \mathbf{Y}_{lm}^{(2)}(\hat n) &= \frac{\mathbf{r}\times \nabla Y_{lm}(\hat n)}{\sqrt{l(l+1)}}.
\end{align}
The magnetization torque $\mathbf{T}$ induced by the applied electric field $\mathbf{E}$ is given by the torquance tensor $\hat K(\hat m)$ which depends on the orientation of the magnetization $\hat m$,
\begin{equation}
    \mathbf{T}=\hat K(\hat m)\mathbf{E},
\end{equation}
and can generally be expanded in the VSH basis as
\begin{equation}
    \hat K(\hat m)=\sum_{lm\nu}\mathbf{Y}_{lm}^{(\nu)}(\hat m)\otimes \mathbf{K}_{lm}^{(\nu)}.
    \label{torqVSH}
\end{equation}
Here $\mathbf{K}_{lm}^{(\nu)}$ are (complex) Cartesian vectors whose structure is determined by the symmetry of the system. In particular, consider the axially symmetric case ($C_{\infty v}$ symmetry) characteristic for a polycrystalline bilayer. In this case, the torquance tensor $\hat K(\hat m)$ should be invariant with respect to the rotation of the crystal around the $z$ axis. Because this rotation does not mix VSH with each other, each term in Eq.\ (\ref{torqVSH}) should be invariant.

If the bilayer is rotated around the $z$ axis by angle $\delta$, the VSH transforms as $\mathbf{Y}_{lm}^{(\nu)}(\theta,\phi)\to \mathbf{Y}_{lm}^{(\nu)}(\theta,\phi-\delta)=e^{-im\delta}\mathbf{Y}_{lm}^{(\nu)}(\theta,\phi)$. Therefore, $\mathbf{K}_{lm}^{(\nu)}$ must transform as $\mathbf{K}_{lm}^{(\nu)}\to e^{im\delta}\mathbf{K}_{lm}^{(\nu)}$, which is only possible for $m=0$ with $\mathbf{K}_{l0}^{(\nu)}\parallel \hat z$, or for $m=\pm1$ with $\mathbf{K}_{l,\pm1}^{(\nu)} =K_{l,\pm1}^{(\nu)} (\hat x\pm i\hat y)$. Thus, only VSH with $m=\pm1$ are allowed in the axially symmetric case for the torque arising in response to the in-plane electric field. We also have $K_{l,-1}^{(\nu)}=-K_{l,1}^{(\nu)*}$, because $\hat K(\hat m)$ must be real.

In this paper we use the reference frame in which $\mathbf{E}=E\hat x$.
Mirror reflection symmetry only allows $\mathrm{Im}\mathbf{Y}_{l1}^{(\nu)}$ for odd $l$ and $\mathrm{Re}\mathbf{Y}_{l1}^{(\nu)}$ for even $l$. It is then convenient to use the basis of (also orthonormal) real VSH, which are defined, similar to the real scalar spherical harmonics, as $\mathbf{Z}_{l,-1}^{(\nu)}=-\sqrt{2}\mathrm{Im}\mathbf{Y}_{l1}^{(\nu)}$ and $\mathbf{Z}_{l,1}^{(\nu)}=-\sqrt{2}\mathrm{Re}\mathbf{Y}_{l1}^{(\nu)}$. The torquance is then represented by the expansion
\begin{equation}
    \mathbf{T}/E=\sum_{l\nu}C^{(\nu)}_{l,(-1)^l}\mathbf{Z}^{(\nu)}_{l,(-1)^l}.
\end{equation}

Apart from being orthonormal, the VSH (both complex and real) have the following useful properties for representing SOT. First, under time reversal ($\hat m\to -\hat m$), $\mathbf{Z}_{lm}^{(1)}$ is even for odd $l$ and odd for even $l$, while the opposite holds for  $\mathbf{Z}_{lm}^{(2)}$. Further, the contribution of a given SOT term to magnetization damping is given by the curl of the effective field $\mathbf{B}=\mathbf{T}\times\hat m$ corresponding to that term. It follows from the definition of VSH that the effective field corresponding to the torque harmonic $\mathbf{Z}_{lm}^{(1)}$ has the form of $\mathbf{Z}_{lm}^{(2)}$, and vice-versa. Because $\mathbf{Z}_{lm}^{(2)}\propto\hat{\mathbf{L}}Z_{lm}$, we find that the torque harmonic $\mathbf{Z}_{lm}^{(1)}$ generates damping proportional to $\hat{\mathbf{L}}^2 Z_{lm}=l(l+1)Z_{lm}$, i.e., simply to $Z_{lm}$. On the other hand, torque harmonics $\mathbf{Z}_{lm}^{(2)}$ do not contribute to damping at all, because the corresponding effective field is a pure gradient and, therefore, has a zero curl. Conversely, the effective field corresponding to torque harmonics $\mathbf{Z}_{lm}^{(1)}$ has zero divergence. These properties make it natural to call $\mathbf{Z}_{lm}^{(1)}$ torque harmonics purely damping-like, and $\mathbf{Z}_{lm}^{(2)}$ purely field-like.

The angular dependence of the current-induced magnetization damping generated by the first two damping-like terms in the VSH expansion, $\mathbf{Z}_{1,-1}^{(1)}$ and $\mathbf{Z}_{2,1}^{(1)}$, is proportional to $m_y$ and $m_xm_z$, respectively \cite{Safranski.Montoya.eaAe2017,Belashchenko2019}.

The commonly used SOT types can be represented in terms of VSH as follows:
\begin{flalign}
&\quad\text{Damping-like:}\quad \hat m\times(\hat y \times \hat m)=\sqrt{8\pi/3}\,\mathbf{Z}_{1,-1}^{(1)}\label{dlvsh}\\
&\quad\text{Field-like:}\quad \hat y \times \hat m=-\sqrt{8\pi/3}\,\mathbf{Z}_{1,-1}^{(2)}\label{flvsh}\\
&\quad\text{Planar-Hall-like \cite{Safranski.Montoya.eaAe2017,Belashchenko2019}:}\nonumber\\
&\qquad m_x \hat m\times(\hat z\times \hat m)=\sqrt{2\pi/3}\,\mathbf{Z}_{1,-1}^{(2)}+\sqrt{2\pi/5}\,\mathbf{Z}_{2,1}^{(1)}&
\label{phvsh}
\end{flalign}
As can be seen from Eqs.\ (\ref{dlvsh})-(\ref{phvsh}), the leading damping-like and field-like SOT terms are pure VSH's, while the planar-Hall-like term (\ref{phvsh}) is a linear combination of the field-like term and a damping-like VSH with $l=2$.
For brevity, we retain the terms \emph{damping-like} and \emph{field-like} for the leading terms (\ref{dlvsh}) and (\ref{flvsh}).

\end{document}